\documentclass[showkeys,showpacs,superscriptaddress,nofootinbib]{revtex4-2}
\usepackage{amsmath}
\usepackage{amssymb}
\usepackage{graphicx}
\usepackage{bm}
\usepackage{float}
\usepackage{ulem,xcolor}
\usepackage{color,soul}
\allowdisplaybreaks
\begin{document}

\title{
	Quasiparticles in nonperturbative vacuum
 }
\date{\today}

\author{
	Vladimir Dzhunushaliev
}
\email{v.dzhunushaliev@gmail.com}
\affiliation{
	Department of Theoretical and Nuclear Physics,  Al-Farabi Kazakh National University, Almaty 050040, Kazakhstan
}
\affiliation{
	Academician J.~Jeenbaev Institute of Physics of the NAS of the Kyrgyz Republic, 265 a, Chui Street, Bishkek 720071, Kyrgyzstan
}

\author{Vladimir Folomeev}
\email{vfolomeev@mail.ru}
\affiliation{
	Academician J.~Jeenbaev Institute of Physics of the NAS of the Kyrgyz Republic, 265 a, Chui Street, Bishkek 720071, Kyrgyzstan
}

\begin{abstract}
The model of nonperturbative vacuum in SU(2) Yang-Mills theory coupled to a nonlinear spinor field is suggested.
By analogy with Abelian magnetic monopole dominance in quantum chromodynamics, it is assumed that the dominant  contribution to
such vacuum is coming from quasiparticles described by dipolelike solutions existing in this theory.
Using an assumption of the behavior of the number density of quasiparticles whose energy approaches infinity, we derive an approximate expression for the energy density
of such nonperturbative vacuum, which turns out to be finite, unlike an infinite energy density of perturbative vacuum.
Using characteristic values of the parameters appearing in the expression for the nonperturbative energy density,
it is shown that this density may be of the order of the energy density associated with Einstein's cosmological constant.
The physical interpretation of the spinor field self-coupling constant as a characteristic distance between quasiparticles is suggested.
The questions of experimental verification of the nonperturbative vacuum model under consideration and of determining its pressure are briefly discussed.
\end{abstract}



\date{\today}

\maketitle

\section{Introduction}

Vacuum is a well-defined concept in quantum theories with free or weakly interacting fields. It consists of virtual particles which are created and annihilated
in accordance with the Heisenberg uncertainty relation. One of basic problems in considering such perturbative vacuum is the presence of an infinite vacuum or zero-point energy,
that in the case of theories which do not involve gravity is artificially removed by subtraction.
%In theories which do not involve gravity, this problem is easily circumvented by using the renormalization procedure.
%subtracting this energy out.

In quantum chromodynamics (QCD), when considering a quark-gluon plasma, one may introduce quasiparticles filling the plasma~\cite{Shuryak:2004tx}.
The quark-gluon plasma is an essentially nonperturbative object where quasiparticles play a quite important role.
Consistent with this, one may assume that in nonperturbative vacuum quasiparticles might also play a quite essential role.
In this connection, the questions can be asked: How such objects may appear in nonperturbative vacuum? How one can describe them?

An attempt to answer these questions will be made in the present paper.
We will consider here a hypothesis according to which, among all possible fluctuations of a quantum field in vacuum,
there are such profiles that satisfy field equations, and they yield a considerable contribution to the corresponding path integral in quantizing such field.
Physically, this means that the properties of such nonperturbative vacuum would be basically described by the aforementioned virtual quasiparticles.
Such quasiparticles appear in accordance with the Heisenberg uncertainty principle and exist over the time
$\Delta t \approx \hbar / \Delta E$ determined by this principle. This idea is analogous to the long-familiar concept that in QCD there exists the so-called
Abelian magnetic monopole dominance. The latter implies that magnetic monopoles, as topologically nontrivial configurations, yield a considerable contribution
to the functional integral. The Abelian and monopole dominance~\cite{Ezawa:1982bf,Ezawa:1982ey} has been confirmed by lattice calculations in QCD (see, e.g., Refs.~\cite{Arasaki:1996sm,Chernodub:1997ay}).

In the present paper we suggest a model of nonperturbative vacuum where monopolelike objects found in Refs.~\cite{Dzhunushaliev:2020qwf,Dzhunushaliev:2021apa}
are regarded as virtual quasiparticles in nonperturbative vacuum of  SU(2) Yang-Mills theory which includes a nonlinear spinor field. As pointed out above, such virtual
quasiparticles are created according to the Heisenberg uncertainty principle and their lifetimes are determined by this principle. Then, if one regards such quasiparticles
as making the main contribution to the nonperturbative vacuum, it is possible to determine such properties of the nonperturbative vacuum like, for example, its energy density.

\section{Dipolelike solutions in SU(2) Yang-Mills theory}

In this section we briefly describe dipolelike solutions in SU(2) Yang-Mills theory with a source of gauge field in the form of a nonlinear spinor field described by the nonlinear Dirac equation.
%We have enclosed the word ``monopole'' in quotation marks because our solution differs in principle from the 't~Hooft-Polyakov monopole solution.
The Lagrangian of such theory is (here we follow Ref.~\cite{Dzhunushaliev:2021apa})
\begin{equation}
\begin{split}
		\mathcal L = & - \frac{1}{4} F^a_{\mu \nu} F^{a \mu \nu}
		+ i \hbar c \bar \psi \gamma^\mu D_\mu \psi  -
		m_f c^2 \bar \psi \psi+
		\frac{l_0^2}{2} \hbar c \left( \bar \psi \psi \right)^2.
\label{1_10}
\end{split}
\end{equation}
Here $m_f$ is the mass of the spinor field;
$
D_\mu = \partial_\mu - i \frac{g}{2} \sigma^a
A^a_\mu
$ is the gauge-covariant derivative, where $g$ is the coupling constant and $\sigma^a$ are the SU(2) generators (the Pauli matrices);
$
F^a_{\mu \nu} = \partial_\mu A^a_\nu - \partial_\nu A^a_\mu +
g \epsilon_{a b c} A^b_\mu A^c_\nu
$ is the field strength tensor for the SU(2) field, where $\epsilon_{a b c}$ (the completely antisymmetric Levi-Civita symbol) are the SU(2) structure constants;
$l_0$ is a constant; $\gamma^\mu$ are the Dirac matrices in the standard representation; $a,b,c=1,2,3$ are color indices and $\mu, \nu = 0, 1, 2, 3$ are spacetime indices.

Unlike Refs.~\cite{Dzhunushaliev:2020qwf,Dzhunushaliev:2021apa,Dzhunushaliev:2022apb}, we will use below the term ``dipole''
for the configurations under consideration. This is because the radial component of the color magnetic field behaves asymptotically as $H^a_r \sim D/r^3$,
as it takes place in Maxwell's magnetostatic, and the constant  $D$ may be called the color magnetic moment of the dipole. But, unlike usual dipole solutions,
the solution obtained in Refs.~\cite{Dzhunushaliev:2020qwf,Dzhunushaliev:2021apa,Dzhunushaliev:2022apb} also has
a nonvanishing component of the color magnetic field $H^a_\varphi$; for this reason, we have enclosed the word ``dipole'' in quotation marks.
Below we discuss a vector field created by the color current $j^{a \mu}$.

Using the Lagrangian~\eqref{1_10}, one can find the corresponding field equations,
\begin{eqnarray}
	D_\nu F^{a \mu \nu} &=& j^{a \mu} =
	\frac{g \hbar c}{2}
	\bar \psi \gamma^\mu \sigma^a \psi ,
\label{1_20}\\
	i \hbar \gamma^\mu D_\mu \psi  - m_f c \psi + l_0^2 \hbar \psi
	\left(
		\bar \psi \psi
	\right)&=& 0 .
\label{1_30}
\end{eqnarray}
Here $j^{a \mu}$ is a color current created by the spinor field $\psi$. ``Dipole'' solutions are sought in the form of the standard {\it Ansatz} used in describing the 't~Hooft-Polyakov monopole,
\begin{equation}
A^a_t = 0, \quad	A^a_i =  \frac{1}{g} \left[ 1 - f(r) \right]
	\begin{pmatrix}
		0 & \phantom{-}\sin \varphi &  \sin \theta \cos \theta \cos \varphi \\
		0 & -\cos \varphi &   \sin \theta \cos \theta \sin \varphi \\
		0 & 0 & - \sin^2 \theta
	\end{pmatrix} , \quad
		i = r, \theta, \varphi  \text{ (in polar coordinates)},
	\label{2_10}
\end{equation}
and the {\it Ansatz} for the spinor field is taken to be~\cite{Li:1982gf,Li:1985gf}
\begin{equation}
	\psi^T = \frac{e^{-i \frac{E t}{\hbar}}}{g r \sqrt{2}}
	\begin{Bmatrix}
		\begin{pmatrix}
			0 \\ - u \\
		\end{pmatrix},
		\begin{pmatrix}
			u \\ 0 \\
		\end{pmatrix},
		\begin{pmatrix}
			i v \sin \theta e^{- i \varphi} \\ - i v \cos \theta \\
		\end{pmatrix},
		\begin{pmatrix}
			- i v \cos \theta \\ - i v \sin \theta e^{i \varphi} \\
		\end{pmatrix}
	\end{Bmatrix},
	\label{2_20}
\end{equation}
where $E/\hbar$ is the spinor frequency and the functions $u$ and $v$ depend on the radial coordinate $r$ only.

Substituting the expressions~\eqref{2_10} and \eqref{2_20} in the field equations~\eqref{1_20} and \eqref{1_30}, one can obtain
equations for the unknown functions $f, u$, and $v$:
\begin{eqnarray}
	- f^{\prime \prime} + \frac{f \left( f^2 - 1 \right) }{x^2} +
	\tilde g^2_{\Lambda} \frac{\tilde u \tilde v}{x} &=& 0 ,
	\label{2_30}\\
	\tilde v' + \frac{f \tilde v}{x} &=& \tilde u \left(
	- 1+ \tilde E +
	\frac{\tilde u^2 - \tilde v^2}{x^2}
	\right) ,
	\label{2_40}\\
	\tilde u' - \frac{f \tilde u}{x} &=& \tilde v \left(
	- 1 - \tilde E +
	\frac{\tilde u^2 - \tilde v^2}{x^2}
	\right),
	\label{2_50}
\end{eqnarray}
%Here, for convenience of making numerical calculations, we have introduced the following dimensionless variables:
written in terms of the following dimensionless variables:
$x = r/\lambda_c$,
$
\tilde u=u\sqrt{l_0^2/\lambda_c g^2},
\tilde v = v\sqrt{l_0^2/\lambda_c g^2},
\tilde E = \lambda_c E/(\hbar c),
\tilde g^2_{\Lambda} = \left( g^\prime \lambda_c/ l_0 \right)^2$,
where $\lambda_c= \hbar / (m_f c)$ is the Compton wavelength and ${g^\prime}^2 = g^2 \hbar c$ is a dimensionless coupling constant. The prime denotes differentiation with respect to  $x$.

\begin{figure}[t]
\begin{minipage}[t]{.49\linewidth}
\centering
	\includegraphics[width=.9\linewidth]{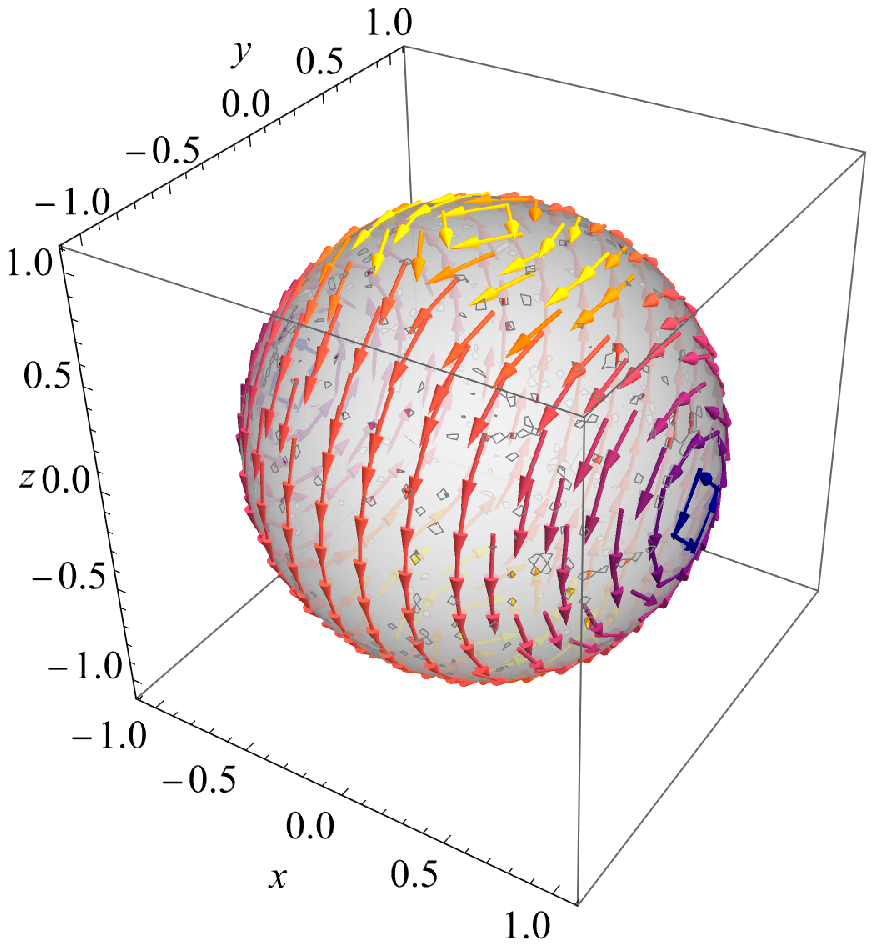}
\end{minipage}\hfill
\begin{minipage}[t]{.49\linewidth}
\centering
	\includegraphics[width=.9\linewidth]{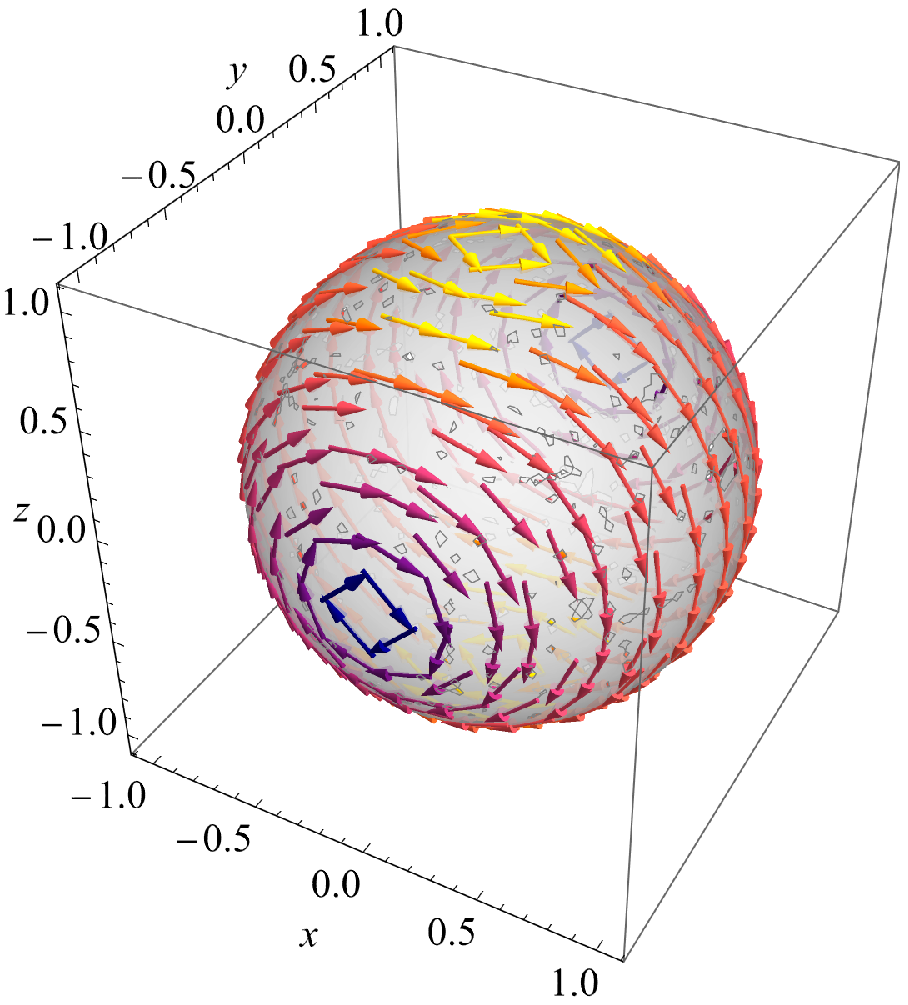}
\end{minipage}
%\vspace{-0.5cm}
\caption{Sketches of the force lines of the components of the color current
 $g \vec{j}^{1} r^3 /(\hbar\, c\,  u v)$ (left panel) and $g \vec{j}^{2} r^3 /(\hbar\, c\,  u v)$ (right panel)
 on the surface $r=\text{const}$. The plots are given in Cartesian coordinates $\{x,y,z\}$.
}
\label{col_cur}
\end{figure}

The color current $j^{a \mu}$ for the spinor \eqref{2_20} has the following components:
$$	\vec{j}^{1} =  \frac{\hbar\, c}{g}\frac{  u v}{ r^3}  \left\lbrace
		0, - \sin \varphi , - \cot \theta \cos \varphi
	\right\rbrace ,
\quad
	\vec{j}^{2} = \frac{\hbar\, c}{g} \frac{ u v}{ r^3}  \left\lbrace
		0, \cos \varphi , - \cot \theta \sin \varphi
	\right\rbrace ,
\quad
	\vec{j}^{3} =  \frac{\hbar\, c}{g}\frac{ u v}{ r^3}  \left\lbrace
		0, 0 , 1
	\right\rbrace ,
$$
where the superscripts $1,2,3$ denote color indices. The current $\vec{j}^{3}$ has only the $\varphi$-component; this means that this is a circular current creating a dipole
with the color index 3. The currents $\vec{j}^{1, 2}$ are shown in Fig.~\ref{col_cur}. It is seen that these currents also form dipoles;
 this suggests that the solution under consideration can be called a ``dipole'' solution, enclosed in quotation marks because it also contains a nonvanishing
 component of the color magnetic field $H^a_\varphi$.

The total energy density of the ``dipole'' under consideration is
\begin{equation}
	\tilde \epsilon =
	\tilde{\epsilon}_m + \tilde \epsilon_s =\frac{1}{\tilde g^2_\Lambda}
	\left[
	\frac{{f'}^2}{ x^2} +
	\frac{\left( f^2 - 1 \right)^2}{2 x^4}
	\right] +
	\left[
	\tilde E \frac{\tilde u^2 + \tilde v^2}{x^2} +
	\frac{\left(\tilde u^2 - \tilde v^2 \right)^2}{2 x^4}
	\right].
	\label{2_60}
\end{equation}
Here the expressions in the square brackets correspond to the dimensionless energy densities of the non-Abelian gauge fields,
$
	\tilde{\epsilon}_m \equiv \frac{\left( l_0 \lambda_c\right)^2}{\hbar c} \epsilon_m
$,
and of the spinor field,
$
	\tilde{\epsilon}_s \equiv \frac{\left( l_0 \lambda_c\right)^2}{\hbar c} \epsilon_s
$.

Correspondingly, the total energy of the ``dipole'' is calculated as
\begin{equation}
	\tilde W_t \equiv \frac{m_f l_0^2}{\hbar^2} W_t =
	4 \pi
	\int\limits_0^\infty x^2 \tilde \epsilon d x
	= \left( \tilde{W}_t \right)_m + \left( \tilde{W}_t \right)_{s},
\label{2_70}
\end{equation}
where the energy density $\tilde \epsilon$ is taken from Eq.~\eqref{2_60}.  In the right-hand side of Eq.~\eqref{2_70}, the first term is the energy of the magnetic field and the second term corresponds to the energy of the nonlinear spinor field.

\begin{figure}[t]
	\centering
	\includegraphics[width=.5\linewidth]{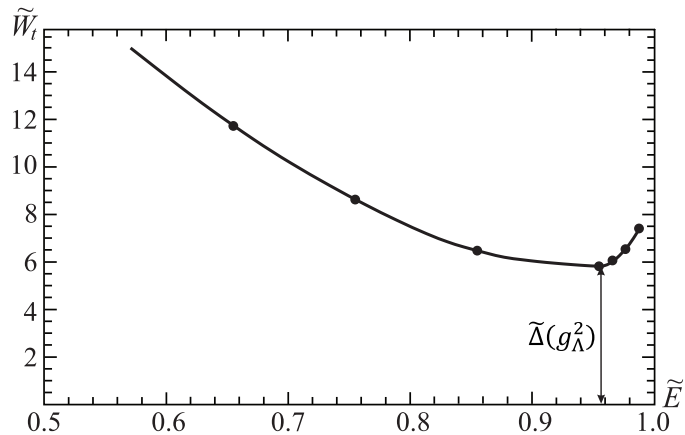}
	%\vspace{-0.5cm}
	\caption{A sketch of the energy spectrum of the total energy from Eq.~\eqref{2_70} as a function of the spinor frequency $\tilde E$
		(for details see Ref.~\cite{Dzhunushaliev:2020qwf}).
	}
	\label{energy_spectrum}
\end{figure}

The typical behavior of  $\tilde W_t $ as a function of the spinor frequency $\tilde{E}$ is sketched in Fig.~\ref{energy_spectrum}.
It is seen that there exists an absolute minimum of the dimensionless energy (mass gap); call it  $\tilde \Delta(g^2_\Lambda)$.
Moreover,  as demonstrated in Refs.~\cite{Dzhunushaliev:2020qwf,Dzhunushaliev:2021apa},  for $\tilde{E}\to 0,1$ the energy of the ``dipole'' goes to infinity.

The asymptotic behavior of the functions $f, \tilde v$, and $\tilde u$ is as follows:
\begin{align}
	f	(x) \approx & 1 - \frac{f_\infty}{x} ,
\label{2_75}\\
	\tilde u(x) \approx & \tilde u_\infty
	e^{- x \sqrt{1 - \tilde E^2}} ,\quad
	\tilde v(x) \approx
	\tilde v_\infty e^{- x \sqrt{1- \tilde E^2}} ,
\label{2_80}
\end{align}
where $f_\infty, \tilde u_\infty$,  and $\tilde v_\infty$ are integration constants.

Consider the question of a linear size of the ``dipole'' under investigation. It consists of two fields~-- the non-Abelian field $A^a_\mu$ and the spinor field $\psi$,
whose asymptotic behavior is fundamentally  different. As follows from Eq.~\eqref{2_75}, the color field decreases according to the power law,
$H^a_{r , \theta, \varphi} \sim r^{-3}$~\cite{Dzhunushaliev:2020qwf}, whereas the spinor field decreases exponentially with distance according to Eq.~\eqref{2_80}.
In this connection, it is difficult to introduce the notion of the size of the ``dipole''.
However, one may introduce the notion of the core of the ``dipole'', which represents the region where the spinor field (which is a source of the color Yang-Mills field)
is concentrated. Then, in accordance with Eq.~\eqref{2_80}, one may define a characteristic size of the core of the ``dipole'' as
%created by the spinor field, which is a source of the color Yang-Mills field, as
\begin{equation}
	l_{cr} \approx \frac{\lambda_c}{\sqrt{1- \tilde E^2}} .
\label{2_90}
\end{equation}
This implies that for $\tilde E \rightarrow 1$ the size of the core $l_{cr} \rightarrow \infty$. In turn, for $\tilde E \rightarrow 0$, it is estimated as $l_{cr} \approx \lambda_c$, i.e., it remains unchanged.
Unfortunately, the power-law asymptotic behavior of the magnetic field \eqref{2_75} does not permit one to define a quantity which adequately represents the size of the ``dipole'' itself, and not just the size of the core.

Taking into account that the linear size of  the core is estimated according to the expression~\eqref{2_90}, a maximal number of quasiparticles (``dipoles'') per unit volume can be estimated as
$$
	\left[ n_{qp} \left( \tilde E\right) \right]_{\text{max}} \approx
	\left[l_{cr}\left( \tilde E\right) \right]^{-3} .
$$
Perhaps such concentration of the quasiparticles is not attainable in vacuum, but it may be reached in a quark-gluon plasma at high temperatures.

\section{``Dipoles'' as quasiparticles in the nonperturbative vacuum
}

In this section we describe a scenario within which ``dipoles'' found in Refs.~\cite{Dzhunushaliev:2020qwf,Dzhunushaliev:2021apa} may be regarded
as quasiparticles in the nonperturbative vacuum of  SU(2) Yang-Mills theory coupled to a nonlinear spinor field.

\begin{figure}[t]
\centering
	\includegraphics[width=.4\linewidth]{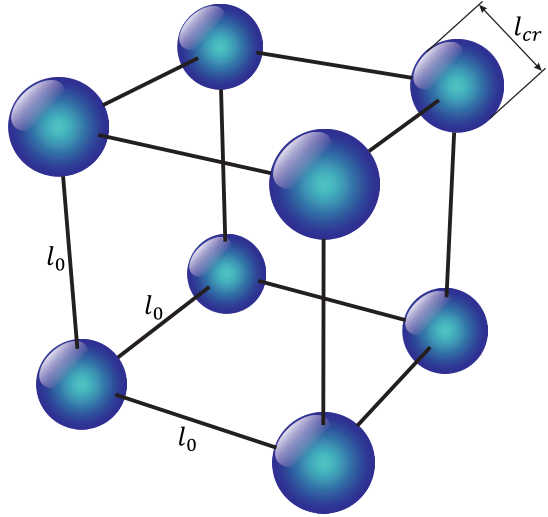}
%	\vspace{-0.5cm}
	\caption{A sketch of the averaged spatial distribution of quasiparticles in the nonperturbative vacuum; their location in space is randomly changed, and this leads to nonzero values of the dispersions
and of higher order Green's functions. The size of the core $l_{cr}$ is characterized by the Compton wavelength of a fermion $\lambda_c$ with the mass $m_f$ [cf. Eq.~\eqref{2_90}].
The average distance between quasiparticles is determined by the spinor field self-coupling constant $l_0$.
		}
\label{QP_in_NP_vacuum}
\end{figure}

The main idea is the same as that used in Abelian magnetic monopole dominance: among all possible fluctuations of a gauge field there are some that satisfy Eqs.~\eqref{2_30}-\eqref{2_50}.
Thus, we suppose that, within the nonperturbative vacuum model under consideration, such configurations
give the dominant contribution to the path integral in the theory with the Lagrangian~\eqref{1_10}.
Since we consider vacuum, the mean vacuum values of all color potentials and of the spinor field are zero,
$$
	\left\langle \hat A^a_\mu \right\rangle =  0 ,
\quad 	\left\langle \hat \psi \right\rangle =  0 ,
$$
whereas the dispersions of these fields are nonvanishing,
$$
	\left\langle \left( \hat A^a_\mu \right)^2 \right\rangle \neq  0 ,
	\quad 	\left\langle \left( \hat \psi\right)^2 \right\rangle \neq  0 .
$$
In the nonperturbative vacuum model under consideration, the dominant contribution to the nonvanishing magnitudes of the dispersions, as well as to higher order Green's functions,
is given by quasiparticles whose averaged spatial distribution is sketched in Fig.~\ref{QP_in_NP_vacuum}.

%The characteristics of the nonperturbative vacuum are essentially determined by
The system under consideration contains two parameters  $\lambda_c$ and $l_0$
having the dimensions of length. First of them determines the size of the core of the ``dipole'' created by the spinor field, and the second one~-- the spinor field self-coupling constant,
which characterizes the distance between quasiparticles (see Fig.~\ref{QP_in_NP_vacuum}).
Thus, the theory contains two constants  with the dimensions of length which determine the properties of the  nonperturbative vacuum:
the size of the core of quasiparticles and the distance between them.

Let $n_{qp}\left( \tilde E \right)$ is the average density of quasiparticles per unit volume, which depends on the spinor frequency  $\tilde E$;
then the average energy density of the nonperturbative vacuum can be defined as
\begin{equation}
	\varepsilon_{0} \approx \frac{\hbar^2}{m_f l_0^2}
	\left\langle n_{qp} \left( \tilde E \right)  \tilde W_t \left(\tilde E\right) \right\rangle ,
\label{3_40_a}
\end{equation}
where the dimensionless energy  $\tilde W_t $ is given by the expression~\eqref{2_70} and
$\left\langle \cdots \right\rangle $ denotes the quantum averaging over all possible values of $\tilde E$.

The lifetime of a virtual quasiparticle with the energy  $W_t \left(\tilde E\right)$ is determined according to the Heisenberg uncertainty principle,
\begin{equation}
	\Delta t \approx \frac{\hbar }{W_t (\tilde E)} =
	\frac{m_f l_0^2}{\hbar} \frac{1}{\tilde W_t (\tilde E)} .
\label{3_45}
\end{equation}
This expression implies that the maximum lifetime is for a quasiparticle with the energy equal to the energy of the mass gap $\tilde W_t\left(\tilde{E}\right) =\tilde \Delta\left(g^2_\Lambda\right)$.

Consider next the behavior of the concentration of quasiparticles $n_{qp} \left( \tilde E \right)$ for
$\tilde{E} \rightarrow 0, 1$. For $\tilde{E} \rightarrow 1$, according to Eq.~\eqref{2_90}, the linear size of the region occupied by the core of the ``dipole'' created by the spinor field
approaches infinity, $l_{cr} \rightarrow \infty$. This actually means that the size of a quasiparticle approaches infinity as well; that is, one quasiparticle occupies all the space.
This in turn means that
\begin{equation}
	n_{qp} \left( \tilde E \rightarrow 1 \right) \rightarrow 0 .
\label{3_46}
\end{equation}
As we pointed out above, for $\tilde{E} \rightarrow 0$, the size of the core is estimated as $l_{cr} \approx \lambda_c$,
whereas the value of the non-Abelian magnetic field goes to infinity and the field fills all the space. As in the case with $\tilde{E} \rightarrow 1$, this means that
\begin{equation}
	n_{qp} \left( \tilde E \rightarrow 0 \right) \rightarrow 0 .
	\label{3_46_a}
\end{equation}
Both conditions \eqref{3_46} and \eqref{3_46_a} can be written as
\begin{equation}
	n_{qp} \left( \tilde E  \right) \rightarrow 0 \text{ when }
	\tilde{E} \rightarrow 0,1 \text{ or, equivalently, when }
	\tilde W_t \left( \tilde E\right) \rightarrow \infty .
\label{3_47}
\end{equation}
Thus, the function $n_{qp} \left( \tilde E\right)$ is equal to zero at the boundaries of the interval
$0 \leqslant \tilde{E} \leqslant 1$;
%, $n_{qp} \left( \tilde E = 0,1 \right) = 0$.
this suggests that for some $\tilde{E}_0$ lying in this interval the concentration  $n_{qp} \left( \tilde E_0 \right)$ has a maximum.

The expression for the energy density \eqref{3_40_a} is the product of two functions: the density of quasiparticles and the dimensionless energy of a quasiparticle.
For $\tilde{E} \rightarrow 0,1$, the first function, according to Eq.~\eqref{3_47}, goes to zero, whereas the second one~-- to infinity.
Consequently, their product in Eq.~\eqref{3_40_a} goes either to zero, or to a constant, or to infinity. In order that the energy density would be finite, it is necessary
to choose the first option. This permits one to estimate the quantity  \eqref{3_40_a} as the product of the characteristic value of the density of quasiparticles
 $n_{qp} \left( \tilde E_{0} \right) $ and the  characteristic value of the energy~-- the mass gap $\tilde\Delta(g^2_\Lambda)$:
\begin{equation}
	\varepsilon_{0} \approx \frac{\hbar^2}{m_f l_0^2}
	n_{qp} \left( \tilde E_{0} \right) \tilde \Delta(g^2_\Lambda) .
\label{3_50}
\end{equation}
This expression indicates that in the model under consideration the energy density of the nonperturbative vacuum  $\varepsilon_{0} $ is a finite quantity, in contrast to
the energy of perturbative vacuum which is infinite.
%provided by an infinite set of null oscillations.
Notice that in these calculations the presence of a mass gap
plays a crucial role. In the absence of the mass gap, the computation procedure is changed considerably and the expression \eqref{3_50} would not already be correct.

One of undetermined quantities in the expression \eqref{3_50} is the concentration of quasiparticles $n_{qp} \left( \tilde E_{0} \right)$
with dimensions $\text{(length)}^{-3}$. In the initial equations \eqref{1_20} and
 \eqref{1_30}, there are two quantities having the dimensions of length; this is the Compton wavelength
   $ \lambda_c $ and the spinor field coupling constant $ l_0$. Therefore, one can assume that, in order of magnitude,  the concentration
  might be estimated as
 \begin{equation}
 	n_{qp} \left( \tilde E_{0} \right) \approx \frac{1}{l_0^{3 - \alpha} \lambda_c^{\alpha}} .
 \label{3_55}
 \end{equation}
The quantity  $\left( l_0^{3 - \alpha} \lambda_c^{\alpha}\right)^{1/3}$ characterizes the distance between quasiparticles.
The simplest case of $\alpha=0$ is sketched in Fig.~\ref{QP_in_NP_vacuum}.

It is of interest to estimate characteristic quantities of the nonperturbative vacuum under consideration. To do so, we start from the assumption that its energy density
\eqref{3_50} is of the order of the energy density associated with Einstein's cosmological constant $\Lambda_{E}$, i.e.,
$\varepsilon_{0}=\varepsilon_{\Lambda_{E}} \approx5.4\times 10^{-9}\text{erg cm}^{-3}$.
%the magnitude of the energy density of nonperturbative vacuum \eqref{3_50}.
As an example, take the following set of parameters: (i)~let $\alpha=0$ in Eq.~\eqref{3_55};
(ii)~choose  $m_f$ equal to, say, the mass of the strange quark $m_s\approx 95\, \text{MeV}$;
(iii)~take the magnitude of the dimensionless mass gap to be $\tilde \Delta \left( g^2_\Lambda \right) \approx  50$~\cite{Dzhunushaliev:2021apa};
(iv)~let $ {g^\prime}^2 \approx 10$ (typical value in QCD). Then we have the following estimates:
\begin{equation}
\label{3_60}
\begin{split}
	&\tilde g^2_\Lambda \approx  2.09 \times 10^{-17} ,
		\; \Delta \left( g^2_\Lambda \right) \approx 1.59 \times 10^{-20} \mathrm{erg}, \;
	m_\Delta \equiv \Delta \left( g^2_\Lambda \right) /c^2 \approx 1.77 \times 10^{-41} \mathrm{g}, \\
		&\lambda_c\approx 2.08\times 10^{-13} \text{cm},\;
l_0 \approx 1.44\times 10^{-4} \text{cm} ,\;
	n_{qp} \approx 3.38 \times 10^{11} \text{cm}^{-3} .
\end{split}
\end{equation}

For some reasons this does not allow us to regard the energy density of the nonperturbative vacuum as that associated with Einstein's cosmological constant. One of these reasons
is that the equation of state corresponding to the cosmological constant is
$
	p_{\Lambda_E} = - \varepsilon_{\Lambda_E} .
$
The determination of the pressure of the nonperturbative vacuum (or its equation of state) is a challenging problem, and we will not address it here.
The point is that the virtual quasiparticles under consideration are extended objects, and it is in particular unclear how they interact with a reflecting wall.
%\textcolor{red}{Another reason is the need for an explanation of the disappearance of the energy density of perturbative vacuum.}

\section{Discussion and conclusions}

In the present paper we have suggested a speculative model of nonperturbative vacuum  in SU(2) Yang-Mills theory coupled to a nonlinear spinor field.
Within this model, by analogy with Abelian magnetic monopole dominance, it is assumed that the main contribution to the energy of the nonperturbative vacuum
comes from quasiparticles, which are here represented by  ``dipoles'' constructed in Refs.~\cite{Dzhunushaliev:2020qwf,Dzhunushaliev:2021apa}.
The remarkable feature of such nonperturbative vacuum is that its energy density is finite.

Summarizing the results,
\begin{itemize}
	\item The model of nonperturbative vacuum in SU(2) Yang-Mills theory coupled to a nonlinear spinor field is suggested, assuming that the main contribution to its characteristics comes from quasiparticles
which are described by the corresponding ``dipole'' solutions.
	\item We present arguments to claim that the number density of quasiparticles whose energy approaches infinity goes to zero.
	\item We have derived the expression for the energy density of the nonperturbative vacuum which, being a sum of energies of quasiparticles, is finite.
	\item Starting from the assumption that the magnitude of the nonperturbative vacuum energy density is of the order of the energy density associated with Einstein's cosmological constant,
we have estimated the characteristic quantities of the nonperturbative vacuum under consideration.
	\item We suggest the physical interpretation of the spinor field self-coupling constant, having the dimensions of length, as a quantity characterizing the distance between quasiparticles in the nonperturbative vacuum.
\end{itemize}

We have pointed out that, although the energy density of the nonperturbative vacuum may be of the order of that associated with Einstein's cosmological constant,
for some reasons it is difficult to identify these two quantities. For example, for the cosmological constant, it is necessary to have a strictly
defined relationship between its energy density and pressure. In further studies on this subject it is necessary to get an expression for the pressure
of  the nonperturbative vacuum which amounts to obtaining an equation of state for such vacuum. The problem of principle one faces here is that the quasiparticles are
(i) quantum and virtual; and (ii) extended objects.

Notice that the SU(2) Yang-Mills theory coupled to a nonlinear spinor field under consideration is non-QCD theory. The difference of principle is that, in QCD,
a spinor field is described by the linear Dirac equation, while here we deal with the nonlinear Dirac equation.
At first glance, the linear and nonlinear Dirac equations are in no way related to each other. However, it is worth mentioning here that
there is a possibility of obtaining the nonlinear Dirac equation as a consequence of nonpertubative quantization in QCD, as discussed in Refs.~\cite{Dzhunushaliev:2020qwf,Dzhunushaliev:2021apa,Dzhunushaliev:2022apb}.

In conclusion, we wish to note that the model of nonperturbative vacuum under consideration is an experimentally testable model, at least in principle.
The lifetime of a virtual quasiparticle  is determined by Eq.~\eqref{3_45}; then, e.g.,
 for the parameters used in obtaining the estimates \eqref{3_60}, one has
$	\Delta t \approx 6.6\times 10^{-8} \mathrm{ sec } $.
In turn, one cubic centimeter contains only $\sim 10^{11}$ virtual quasiparticles. Since this concentration and the lifetimes are quite large quantities,
one may expect that an experimental verification of the existence of such virtual quasiparticles might be possible.
%an extremely small quantity,
%run into great difficulty.

\section*{Acknowledgments}

This research has been funded by the Science Committee of the Ministry of Education and Science of the Republic of Kazakhstan.
We are also grateful to the Research Group Linkage Programme of the Alexander von Humboldt Foundation for the support of this research.


\begin{thebibliography}{99}
	
\bibitem{Shuryak:2004tx}
E.~V.~Shuryak and I.~Zahed,
Towards a theory of binary bound states in the quark gluon plasma,
Phys.\ Rev.\ D {\bf 70}, 054507 (2004).

%\cite{Ezawa:1982bf}
\bibitem{Ezawa:1982bf}
Z.~F.~Ezawa and A.~Iwazaki,
Abelian Dominance and Quark Confinement in Yang-Mills Theories,
Phys. Rev. D \textbf{25}, 2681 (1982).

%\cite{Ezawa:1982ey}
\bibitem{Ezawa:1982ey}
Z.~F.~Ezawa and A.~Iwazaki,
Abelian Dominance and Quark Confinement in {Yang-Mills} Theories. 2. Oblique Confinement and $\eta^\prime$ Mass,
Phys. Rev. D \textbf{26}, 631 (1982).

%\cite{Arasaki:1996sm}
\bibitem{Arasaki:1996sm}
N.~Arasaki, S.~Ejiri, S.~i.~Kitahara, Y.~Matsubara and T.~Suzuki,
Monopole action and monopole condensation in SU(3) lattice QCD,
Phys. Lett. B \textbf{395}, 275 (1997).

%\cite{Chernodub:1997ay}
\bibitem{Chernodub:1997ay}
M.~N.~Chernodub and M.~I.~Polikarpov,
Abelian projections and monopoles,
Contribution to: NATO Advanced Study Institute on Confinement, Duality and Nonperturbative Aspects of QCD,
[arXiv:hep-th/9710205 [hep-th]].
%160 citations counted in INSPIRE as of 12 Apr 2022

%\cite{Dzhunushaliev:2020qwf}
\bibitem{Dzhunushaliev:2020qwf}
V.~Dzhunushaliev, V.~Folomeev and A.~Serikbolova,
Monopole solutions in SU(2) Yang-Mills-plus-massive-nonlinear-spinor-field theory,
Phys. Lett. B \textbf{806}, 135480 (2020).
%5 citations counted in INSPIRE as of 12 Apr 2022

%\cite{Dzhunushaliev:2021apa}
\bibitem{Dzhunushaliev:2021apa}
V.~Dzhunushaliev, N.~Burtebayev, V.~N.~Folomeev, J.~Kunz, A.~Serikbolova and A.~Tlemisov,
Mass gap for a monopole interacting with a nonlinear spinor field,
Phys. Rev. D \textbf{104}, 056010 (2021).
%3 citations counted in INSPIRE as of 12 Apr 2022

%\cite{Dzhunushaliev:2022apb}
\bibitem{Dzhunushaliev:2022apb}
V.~Dzhunushaliev and V.~Folomeev,
QCD effects in non-QCD theories,
[arXiv:2203.06851 [hep-ph]].
%0 citations counted in INSPIRE as of 17 Apr 2022

\bibitem{Li:1982gf}
X.~z.~Li, K.~l.~Wang, and J.~z.~Zhang,
Light Spinor Monopole,
Nuovo Cim.\ A {\bf 75}, 87 (1983).

\bibitem{Li:1985gf}
K.~L.~Wang and J.~Z.~Zhang,
The Problem of Existence for the Fermion-Dyon Selfconsistent Coupling System in a SU(2) Gauge Model,
Nuovo Cim.\ A {\bf 86}, 32 (1985).



\end{thebibliography}
\end{document}